# Decoupling structural and bonding effects on ferroelectric switching in ScAlN via molecular dynamics under an applied electric field

[1]Ryotaro Sahashi, [1]Po-Yen Chen, and [2]Teruyasu Mizoguchi
[1]Department of Materials Engineering, The University of Tokyo, Tokyo, Japan
[2]Institute of Industrial Science, The University of Tokyo, Tokyo, Japan

**Abstract**
$Sc_xAl_{1-x}N$ has emerged as a promising wurtzite-type ferroelectric material, where increasing the Sc composition reduces both the coercive field ($E_c$) and remanent polarization ($P_r$). This composition-dependent behavior is physically attributed to two simultaneous changes: the increase in the internal structural parameter $u$ (structural effect) and the weakening of bond strength (bonding effect). Because these factors are strongly coupled in experiments, their individual contributions to ferroelectric switching remain unclear. In this study, we systematically decoupled these effects using machine-learning force field-based molecular dynamics (MD) simulations under an applied electric field. By artificially tuning $u$ via in-plane strain at a fixed composition, we demonstrated that $P_r$ is determined exclusively by the structural effect, exhibiting a universal linear dependence regardless of the composition. In contrast, $E_c$ deviated from this structural trend, implying an additional compositional contribution. To isolate this, we evaluated configurations with identical $u$ but varying Sc compositions; $P_r$ remained constant, whereas $E_c$ systematically decreased due to bond weakening. Furthermore, static nudged elastic band (NEB) calculations revealed that the static switching barrier depends solely on $u$, failing to explicitly capture the bonding effect on $E_c$. These results establish that while $P_r$ is governed strictly by the structural effect, $E_c$ is determined by a superposition of structural and bonding effects. Our findings highlight the necessity of dynamic MD simulations for fully understanding ferroelectric switching in compositionally tunable materials.

## 1. Introduction

Wurtzite-type ferroelectrics, particularly $Sc_xAl_{1-x}N$, have attracted significant attention for applications in next-generation non-volatile memory and neuromorphic devices due to their high compatibility with complementary metal-oxide-semiconductor (CMOS) processes. [1–5] While pure AlN exhibits a coercive field ($E_c$) that is too high to allow conventional polarization switching, the introduction of Sc effectively lowers $E_c$, leading to the emergence of clear ferroelectricity. [6–8] However, this is accompanied by a trade-off: increasing the Sc composition simultaneously reduces the remanent polarization ($P_r$). [9–11] From the perspective of device applications, it is essential to accurately understand and control this strong composition dependence of the ferroelectric properties ($E_c$ and $P_r$). [12]

Microscopically, the degradation of ferroelectric properties upon Sc incorporation is theoretically attributed to two primary physical factors. The first is the structural effect. The substitution of Sc increases the internal structural parameter $u$ (i.e., flattening the wurtzite structure), which shortens the required atomic displacement during polarization switching.[13–15] The second is the bonding effect. Compared to Al-N bonds, Sc-N bonds exhibit weaker covalency (higher ionicity), and this weakening of bond strength inherently affects the energy barrier required for polarization switching.[16–18] However, in actual experimental systems, varying the Sc composition inevitably induces both the increase in the $u$ parameter (structural effect) and the weakening of bond strength (bonding effect) simultaneously. Consequently, it has been experimentally impossible to isolate and quantify the individual contributions of each factor to the reduction of $E_c$ and $P_r$.

Decoupling these fundamental structural and bonding effects is of paramount importance because it provides a critical pathway to independently control the macroscopic properties, $P_r$ and $E_c$. In practical applications, achieving a low $E_c$ while maintaining a robust $P_r$ is the ultimate goal; a high $P_r$ ensures strong signal readability and reliable data retention, whereas a low $E_c$ is indispensable for low-power, low-voltage CMOS-compatible operations. By fundamentally understanding which microscopic factor governs $P_r$ and which governs $E_c$, it becomes possible to break the intrinsic trade-off.

Furthermore, from a computational perspective, many previous studies have relied on static analyses, such as the nudged elastic band (NEB) method.[19–21] Because these static approaches are limited to evaluating the energy barrier along a hypothetical, homogeneous switching pathway, there is a fundamental concern that they fail to fully capture how the bonding effect influences $E_c$ during the dynamic polarization switching process—such as domain wall formation and dynamic bond breaking/forming—under an actual applied electric field. To bypass the intrinsic trade-off between the prohibitive computational cost of first-principles calculations and the limited predictive power of empirical potentials, machine-learning force fields (MLFFs) have recently emerged as a transformative tool in materials modeling.[22–25] By leveraging sophisticated architectures—ranging from equivariant graph neural networks such as MACE, ORB, and SevenNet [26–28] to descriptor-based models like GRACE—modern MLFFs map complex local atomic environments to ab initio energies and forces with unprecedented fidelity. Consequently, they enable large-scale, finite-temperature molecular dynamics (MD) without sacrificing DFT-level accuracy. This data-driven paradigm has already proven highly successful in capturing complex phase transitions and dynamic structural evolutions in prototypical ferroelectrics such as $BaTiO_3$,[29] making it an ideal and robust approach for directly simulating the composition-dependent dynamic switching in wurtzite systems.

Leveraging this advanced computational framework, we employed MLFF-based MD simulations under an applied electric field to address the challenges in ScAlN. Taking full advantage of the flexibility of computational simulations, we artificially constructed two distinct conditions to decouple these strongly correlated factors: (1) modulating only the $u$ parameter via in-plane strain while keeping the composition fixed (to extract the structural effect), and (2) varying the composition while keeping the $u$ parameter fixed (to extract the bonding effect). This approach enabled us to completely separate and independently evaluate the contributions of the structural effect and the bonding effect to both $E_c$ and $P_r$. In this paper, we successfully reproduce the composition-dependent changes in the ferroelectric properties of ScAlN using MD simulations under an applied electric field and then apply the decoupling methodology. Our results explicitly reveal that the decrease in $P_r$ can be entirely explained by the structural effect originating from the internal parameter $u$. In contrast, the decrease in $E_c$ is governed by a superposition of both the structural effect and the bonding effect. Furthermore, through a systematic comparison with static NEB calculations, we demonstrate that the NEB method predominantly captures the structural effect, thereby underestimating the $E_c$ reduction caused by the change in bond strength (the bonding effect). These findings not only provide a comprehensive understanding of the polarization switching mechanisms in ScAlN but also strongly highlight the indispensability of dynamic analysis (MD)—over conventional static methods (NEB)—for accurately predicting the switching behaviors of compositionally tunable ferroelectrics.

## 2. Results and Discussion

To investigate the ferroelectric switching behavior of $Sc_xAl_{1-x}N$ under an applied electric field, MD simulations were performed by combining MLFF-predicted interatomic forces with BEC-derived electric-field-induced forces. The interatomic interactions were described using a fine-tuned MACE model trained on a publicly available first-principles dataset.[30] Meanwhile, the BEC tensors required for external field coupling were predicted using the equivar_eval model. This model was trained on a custom BEC dataset constructed in our preceding work using the structural configurations from the the first-principles dataset described above.[31] Both models, whose accuracy and reliability were previously established,[31] were directly adopted in the present simulations. The external electric-field-induced force acting on atom $i$ was evaluated at each MD step following the approach proposed by Chen et al.[32,33] according to Equation 1:

$$F_{\text{ext}} = |e| E_\beta Z^*_{i,\beta\alpha} \quad (1)$$

where $e$ is the elementary charge, $E_\beta$ is the applied electric-field vector, and $Z^*_{i,\beta\alpha}$ is the BEC tensor predicted by the equivar_eval model. The total force acting on each atom during the MD time integration was obtained as the sum of the interatomic force predicted by the MACE model and the external electric-field-induced force defined above. To mimic the epitaxial constraint (substrate clamping effect) typically present in thin-film ScAlN and to better disentangle the contributions of structural and chemical effects, the present setup fixes the in-plane lattice parameters ($a$ and $b$). All MD simulations under an applied electric field were performed under the NPT ensemble. The electric field was applied in a stepwise manner along the $c$-axis, with a field increment of 1 MV/cm at each step, and was swept cyclically within the range of −40 to +40 MV/cm. Using this simulation framework, the $P$–$E$ hysteresis behavior of $Sc_xAl_{1-x}N$ was systematically analyzed to clarify the roles of structural distortion and compositional effects in determining the ferroelectric switching properties.

## 2.1. Reproduction of the Composition Dependence of Ferroelectric Properties

To investigate the microscopic mechanisms governing the ferroelectric switching in $Sc_xAl_{1-x}N$, we first performed MD simulations under an applied electric field using the MLFF. To ensure the reliability of the switching behavior, we adopted simulation parameters—including the supercell size and the short-range interaction model—that were validated in our previous study.[31] These established conditions confirm that finite-size effects are negligible in our present setup.

**Figure 1**(a) shows the calculated $P$–$E$ hysteresis loops for $Sc_xAl_{1-x}N$ at varying Sc compositions ($x$ = 0.125, 0.25, and 0.375). The dynamic switching behavior is clearly captured in our simulations. As extracted from the $P$–$E$ loops, the $P_r$ decreases from 111 to 98 μC/cm$^2$, and the $E_c$ significantly drops from 22.44 to 11.11 MV/cm with increasing Sc concentration from $x$ = 0.125 to 0.375 (Figure 1(b), (c)). This overall trend is in excellent qualitative agreement with previous experimental reports. In these simulations, fixing the $a$ and $b$ axes inherently mimics the substrate clamping effect, which restricts the dynamic lateral relaxation of the lattice during polarization reversal and appropriately reflects the switching energy barrier characteristic of thin-film ScAlN.

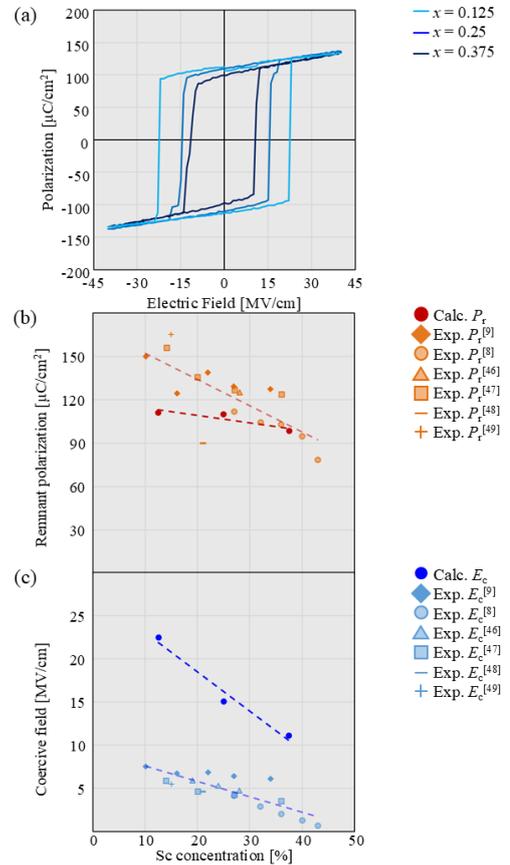

**Figure 1.** Composition dependence of ferroelectric properties in $Sc_xAl_{1-x}N$. (a) Calculated $P$–$E$ hysteresis loops for $x$ = 0.125, 0.25, and 0.375 obtained from MD simulations under an applied electric field. The curves are shown in light blue ($x$ = 0.125), blue ($x$ = 0.25), and navy ($x$ = 0.375), respectively. (b) $P_r$ as a function of Sc concentration. Experimental data are included from Ref.[8,9,50–53]. (c) $E_c$ as a function of Sc concentration. Symbols represent experimental values reported in Ref. [8,9,50–53], showing a systematic reduction of $E_c$ with increasing Sc concentration.

However, this composition-induced degradation inherently involves two simultaneous changes: the flattening of the wurtzite structure (i.e., an increase in the internal structural parameter $u$) and the weakening of the chemical bonds. As explicitly detailed in Supplementary **Figure S1**, our structural models confirm that an increase in the Sc composition naturally drives both the increase in $u$ and the decrease in the average bond strength (-ICOHP) at the same time. Consequently, it is impossible to determine from this compositional variation alone which underlying physical factor fundamentally dictates the reduction of $P_r$ and $E_c$. In the following sections, we independently investigate the structural and chemical bond factors to clarify their respective effects on the ferroelectric properties.

## 2.2. Isolation of the Structural Effect via In-Plane Strain

To extract the purely structural contribution—referred to herein as the structural effect—we artificially modulated the $u$ parameter by applying the in-plane strain from -3% to 3%. Since we found the $u$ parameter has a positive linear relation with in-plane strain as shown in **Figure S2**, the resulting u parameters with a fixed composition of $x = 0.25$ linearly increase from 0.372 to 0.404. As shown in **Figure 2**(a), increasing the $u$ parameter results in a continuous reduction in both $P_r$ and $E_c$.

To quantitatively decouple the structural and bonding contributions, we then plotted $P_r$ and $E_c$ as functions of the $u$ parameter for both the composition-varied and the strain-varied (purely structural) datasets (Figure 2(b), (c)). Remarkably, in the case of $P_r$, the data points from both conditions broadly overlap and follow a highly consistent overall trend (Figure 2(b)). When comparing structures with identical $u$ parameters, the evaluated $P_r$ values remain fundamentally comparable regardless of the varying Sc composition. This strong structural correspondence clearly indicates that the reduction in $P_r$ is primarily governed by the structural effect alone.

In stark contrast, a distinct deviation between the two conditions emerges in the $E_c$ plot (Figure 2(c)). While the purely structural modulation induces a moderate and gradual reduction in $E_c$, the conventional compositional variation drives a significantly more drastic drop with respect to the $u$ parameter. This clear divergence demonstrates that the reduction in $E_c$ cannot be explained solely by the structural flattening; rather, an additional bonding effect must be acting cooperatively to further lower the $E_c$.

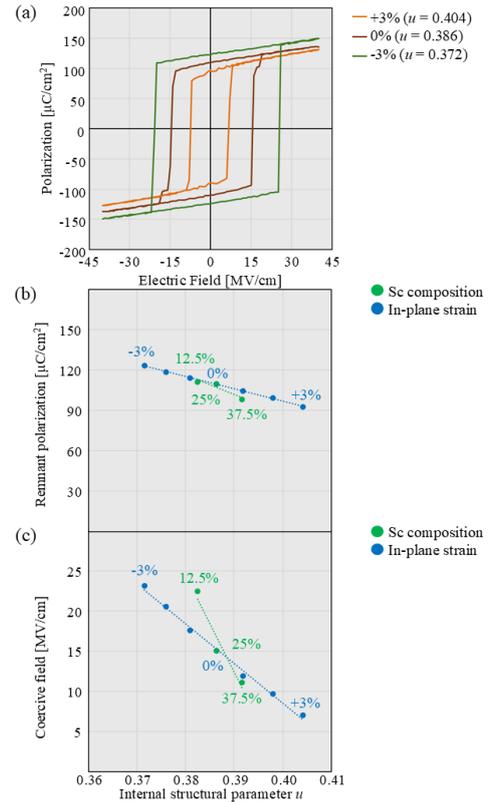

**Figure 2.** Structural control of $P_r$ and $E_c$ via in-plane strain. (a) $P$–$E$ hysteresis loops for $x = 0.25$ under −3%, 0%, and +3% in-plane strain. The curves are shown in green (−3%), brown (0%), and orange (+3%), respectively. (b) $P_r$ as a function of the internal structural parameter $u$. Blue symbols denote structural-control calculations at fixed composition ($x = 0.25$) obtained by in-plane strain, while green symbols correspond to the composition-dependent results shown in Figure 1. (c) $E_c$ as a function of $u$. Blue symbols denote structural-control calculations at fixed composition ($x = 0.25$), and green symbols correspond to the composition-dependent results shown in Figure 1.

## 2.3. Extraction of the Bonding Effect under Fixed Structural Parameters

In addition to structural factors, compositional variation mainly affects the material through modifications in chemical bonding. To investigate the effect of chemical bonding on the ferroelectric properties, we performed an integrated crystal orbital Hamilton population (ICOHP) analysis to quantify the bond strength for the composition-varied and in-plane strain-varied data, where larger −ICOHP value correspond to stronger bonding interactions as shown in **Figure S3**. "It can be clearly observed that larger −ICOHP values correlate with higher $P_r$ and $E_c$, regardless of whether the composition or the in-plane strain is varied. The deviation between these two relationships indicates an additional compositional contribution to the ferroelectric properties beyond the structural effect.

To provide direct computational proof that the observed deviation in $E_c$ originates from the bonding effect, we constructed a specialized set of structural models where the Sc composition varies, but the internal structural parameter is artificially constrained to a constant value ($u \approx 0.386$). First, using crystal orbital Hamilton population (-ICOHP) analysis, we quantitatively confirmed the bond weakening: as the Sc composition increases, the average -ICOHP value decreases from 4.46 to 4.04 eV (**Figure 3**).

Subsequently, we performed MD simulations under an applied electric field on these fixed-$u$ structures. Consistent with our earlier deduction that the $P_r$ is primarily governed by the structural effect, $P_r$ remained remarkably constant at approximately 109 μC/cm² regardless of the decreasing -ICOHP values associated with higher Sc compositions (**Figure 4**(a)). In contrast, even with the structural parameter $u$ tightly constrained, $E_c$ exhibited a clear decrease from 18.54 to 13.82 MV/cm in direct response to the weakening of the bond strength (**Figure 4**(b) and (c)).

These findings provide definitive evidence that while the degradation of $P_r$ is purely a structural phenomenon, the reduction of $E_c$ is determined by a superposition of both the structural effect and the bonding effect. This exact decoupling reveals the underlying physical origins of the macroscopic ferroelectric responses. The purely structural dependence of $P_r$ is physically reasonable; in wurtzite nitrides, the spontaneous polarization is fundamentally a geometric consequence of the symmetry breaking dictated by the internal parameter $u$, consistent with previous DFT investigations.[34,35] Changing the chemical bond strength without altering the atomic positions does not change the macroscopic dipole moment.

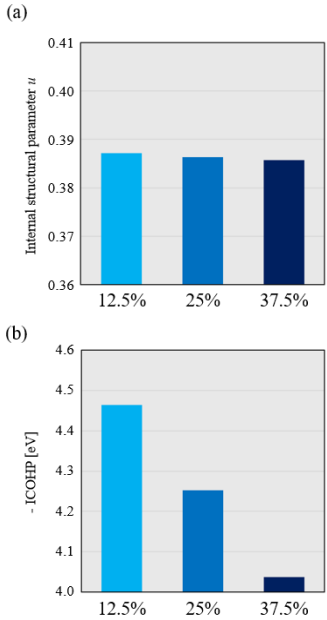

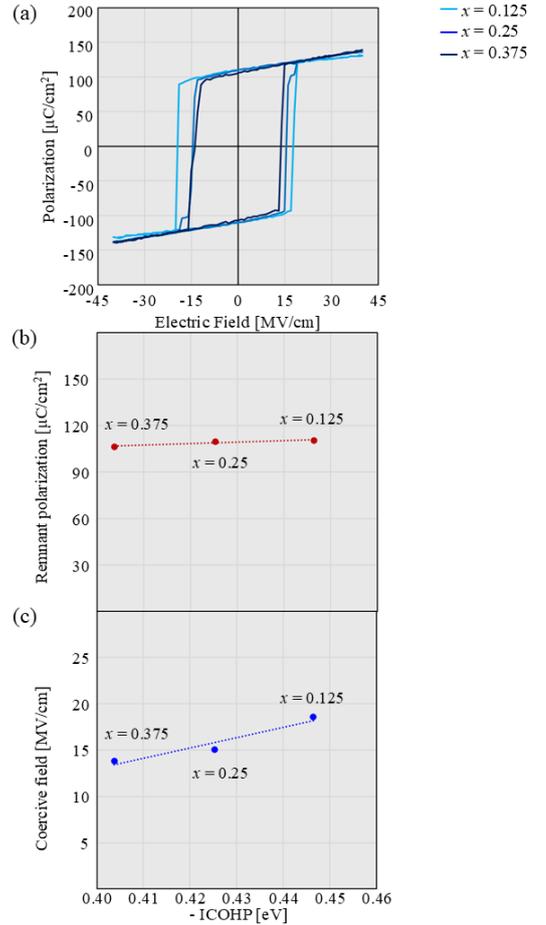

**Figure 3.** Structural matching and bonding variation across compositions. (a) Internal structural parameter $u$ for configurations constructed to achieve nearly identical structural distortion across different Sc concentrations. (b) Corresponding bond-strength metrics quantified via −ICOHP analysis, showing systematic bond weakening with increasing Sc content despite matched structural parameters.

**Figure 4.** Bond–structure decoupling under fixed structural distortion. (a) $P$–$E$ hysteresis loops for $x = 0.125$, 0.25, and 0.375 under nearly identical internal structural parameter $u$. (b) $P_r$ under fixed structural distortion, showing minimal compositional variation. (c) Correlation between bond strength (−ICOHP) and $E_c$, demonstrating systematic reduction of switching field with bond weakening even under fixed structural conditions.

On the other hand, the reduction in $E_c$ requires overcoming a dynamic switching energy barrier. Previous DFT studies have suggested that Sc alloying increases the ionic character and softens the chemical bonds,[36,37] flattening the macroscopic energy landscape. Our decoupled MD simulations explicitly demonstrate that while the structural flattening geometrically reduces the required atomic displacement to reach the transition state, the concurrent bond softening dynamically lowers the resistance to lattice deformation under an external field. Thus, the $E_c$ reduction is dictated by the synergistic superposition of both the structural proximity to the transition state and the chemical flexibility.

## 2.4. Limitations of Static Analysis: A Comparative Study with the NEB Method

Finally, to highlight the necessity of employing dynamic simulations, we compared our findings with static energy barrier calculations using the NEB method (**Figure 5**). As visually illustrated by the representative atomic structures along the switching pathway (**Figure S4**), the transition proceeds mainly through the displacement of cations along the polar axis while maintaining the wurtzite-derived framework. In conventional composition-varied models (where both $u$ and bond strength change), the intrinsic energy barrier for polarization switching decreased substantially from 2.58 to 1.89 meV/f.u as the Sc concentration increases. However, when we evaluated the fixed-$u$ models—where only the bond strength weakens—the static barrier height remained almost unchanged at approximately 2.20 meV/f.u.

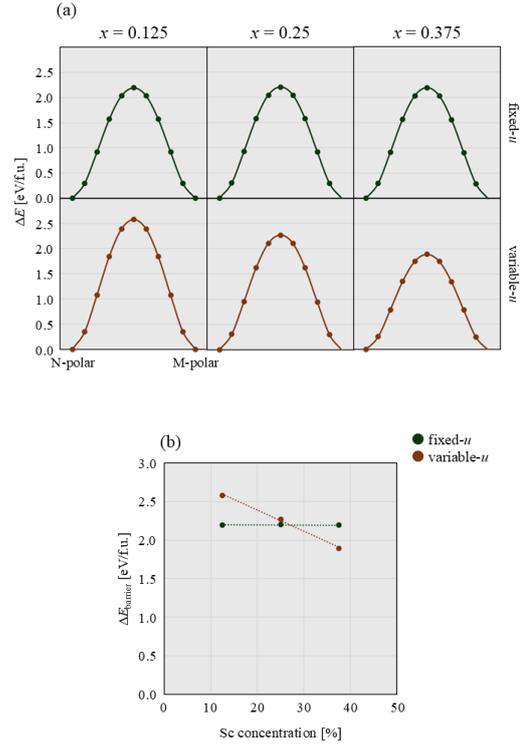

**Figure 5.** Static switching barriers from NEB calculations under different structural constraints. (a) Minimum-energy paths obtained from NEB calculations for $x$ = 0.125, 0.25, and 0.375. The upper row corresponds to configurations with fixed in-plane lattice parameters (matched internal structural parameter $u$), while the lower row corresponds to variable-cell conditions allowing $u$ to relax. The horizontal axis represents NEB image index and the vertical axis denotes relative energy. Barrier differences are negligible under matched structural distortion but become substantial when $u$ differs. (b) Extracted barrier energies as a function of composition for fixed-$u$ and variable-$u$ conditions. The results demonstrate that static barrier calculations primarily reflect structural distortion rather than compositional bonding effects.

This static result stands in stark contrast to the dynamic switching behavior observed in our MD simulations. As explicitly demonstrated in Figure 4(c), the $E_c$ decreases significantly in response to bond weakening, even when the structural parameter $u$ is strictly fixed. In contrast, the NEB method predicts almost no change in the intrinsic energy barrier for these exact same fixed-$u$ configurations. This crucial discrepancy implies that static NEB calculations are predominantly sensitive to the structural effect ($u$ parameter) but fundamentally fail to capture the dynamic influence of bond weakening on the switching field.

Therefore, our results strongly suggest that employing dynamic, MD simulations under an applied electric field is an absolute necessity for accurately predicting the $E_c$ in compositionally tunable ferroelectrics.

## 3. Conclusion

In conclusion, we successfully decoupled the structural and compositional contributions to the degradation of ferroelectric properties in Sc$_x$Al$_{1-x}$N using MLFF-based MD simulations under an applied electric field. Because increasing the Sc composition simultaneously induces structural flattening (an increase in the internal parameter $u$) and the weakening of chemical bonds, these strongly correlated effects have historically been impossible to isolate in experiments.

By artificially modulating the structure via in-plane strain and constraining the $u$ parameter across different compositions, we unambiguously identified the distinct physical origins governing the switching characteristics. We demonstrated that the reduction in $P_r$ is exclusively determined by the purely structural effect (the $u$ parameter), independent of the chemical bond strength. Physically, this confirms that the spontaneous macroscopic polarization is governed geometrically by the atomic coordinates, irrespective of the underlying chemical bond strength. In contrast, the reduction in the $E_c$ cannot be explained by structural changes alone; rather, it is dictated by a superposition of both the structural effect and the bonding effect. Namely, $E_c$ is lowered not only because the flattened structure is geometrically closer to the transition state, but also because the weakened chemical bonds dynamically reduce the resistance to field-induced lattice deformation.

Furthermore, our comparative study with the static NEB method revealed a critical methodological insight: static energy barrier calculations predominantly capture the structural effect but fundamentally fail to reflect the dynamic influence of bond weakening on $E_c$. This highlights that dynamic, MD simulations under an applied electric field are strictly indispensable for accurately predicting and understanding the $E_c$ in compositionally tunable ferroelectrics.

Ultimately, the physical insights and the computational decoupling strategy presented in this study not only clarify the complex switching mechanism of ScAlN but also provide a robust foundation for the rational design and precise property engineering of emerging wurtzite-type ferroelectrics and related functional materials.

## 4. Methods

### 4.1. Machine-Learning Force Field

A MLFF based on the MACE framework was employed to model Sc$_x$Al$_{1-x}$N. The general-purpose MACE-MP-0 model[38] was fine-tuned using a publicly available first-principles dataset reported in Ref.[30]. The reference calculations were performed using VASP within the PAW framework and the PBE exchange–correlation functional.[39–44] The dataset consists of 1664 structures based on 256-atom supercells at compositions $x$ = 0, 0.125, 0.25, and 0.375. Total energies, atomic forces, and stress tensors were used for training. Model architecture and training procedures follow Ref.[31].

Unless otherwise noted, all MD and NEB calculations were performed using the fine-tuned MACE model.

### 4.2. Born Effective Charge Evaluation

To incorporate electric-field effects, BECs were evaluated using the equivariant graph neural network model equivar_eval (BM1 architecture)[45] following Ref.[31]. Reference BEC values were computed using density functional perturbation theory (DFPT) within VASP employing the PBE functional for relaxed structures. The trained BEC model provides a framework for efficiently evaluating BEC tensors corresponding to atomic configurations, enabling quantitative estimation of polarization and electric-field-induced forces beyond purely structural information.[46–48]

### 4.3. MD simulations under an applied electric field

MD simulations under an applied electric field were performed using 360-atom supercells under three-dimensional periodic boundary conditions. Simulations were conducted in the NPT ensemble at 250 K, with a time step of 1 fs for the integration of the equations of motion. The in-plane lattice parameters (*a* and *b* axes) were fixed while allowing relaxation along the *c* axis. This constraint reflects the typical thin-film boundary condition in which in-plane strain is imposed by the substrate, while out-of-plane lattice relaxation remains accessible.

An external electric field was applied along the polarization direction using a triangular waveform to construct full *P–E* hysteresis loops. The field sweep rate was 0.05 kV/(cm·fs). This rate was chosen to balance computational feasibility with quasi-static switching behavior and was verified to produce stable hysteresis loops without artificial dynamic overshoot.

At each electric-field increment, the system was equilibrated for 20 ps, followed by 10 ps of sampling to evaluate polarization. The total simulation time per half-cycle exceeded the characteristic lattice relaxation time, ensuring convergence of the switching response. Polarization was calculated according to Equation 2 as

$$P = \frac{e}{V}\sum_{i} Z_i^* \Delta z_i \qquad (2)$$

where $Z_i^*$ are the ML-predicted BECs and $\Delta z_i$ are atomic displacements relative to a centrosymmetric nonpolar reference structure.

The validity of our simulation setup, including the chosen supercell size and the default short-range interaction cutoff, has been established in a previous study[31]. Prior investigations confirmed that finite-size effects are negligible for this system size, and that extending the interactions to include long-range electrostatics (LES)[49] does not significantly alter the overall switching behavior. Therefore, the main trends reported in the present work are robust with respect to both the system size and the treatment of long-range interactions.

### 4.4. Structural Control via In-Plane Strain

To isolate structural effects, in-plane strain was applied while fixing the a and b lattice parameters and allowing relaxation along the c axis, thereby tuning the internal structural parameter *u*. Switching properties were evaluated using the same MD simulations under an applied electric field protocol.

The structural distortion of the wurtzite lattice was characterized using the internal structural parameter *u*, which describes the relative displacement between the cation and anion sublattices along the *c*-axis. Because polarization switching in wurtzite ferroelectrics occurs primarily through atomic displacements along the *c*-axis, this parameter provides a convenient measure of the structural state relevant to ferroelectric switching.

The internal parameter *u* is defined according to Equation 3 as

$$u = \frac{a^2}{3c^2} + 0.25 \qquad (3)$$

where *a* and *c* denote the lattice constants of the hexagonal unit cell. Variations in *u* correspond to changes in the relative atomic positions along the *c*-axis and therefore directly affect the polarization switching characteristics.

### 4.5. Nudged Elastic Band Calculations

Switching energy barriers were evaluated using the NEB method implemented with the fine-tuned MACE model. Nine intermediate images (eleven images in total) were generated between the initial and final polarization states. The initial and final structures correspond to the N-polar and M-polar configurations, respectively. These structures were obtained from 0 MV/cm configurations extracted from MD simulations under an applied electric field and subsequently relaxed to local minima prior to NEB interpolation.

Two types of structural conditions were considered: (1) configurations with matched internal structural parameter *u*, achieved by fixing the in-plane lattice constants, and (2) configurations in which *u* was allowed to vary. This distinction enabled direct evaluation of structural contributions to the switching barrier. Forces were converged to $F_{max}$ = 0.01 eV/Å. Barrier energies were obtained from the maximum energy along the minimum-energy path relative to the initial N-polar structure, which was used as the reference for the energy difference (Δ*E*).

All NEB calculations were performed under periodic boundary conditions using the same supercell size as the MD simulations. The climbing-image scheme was employed to accurately locate the transition state along the minimum-energy pathway.

**4.6. Bonding Analysis**
To quantify compositional effects on bonding −ICOHP values were calculated using the LOBSTER code based on density functional theory (PBE) recalculations of representative structures. The wavefunctions required for LOBSTER analysis were obtained from VASP calculations using the same exchange–correlation functional as the training dataset to ensure consistency.

Bond-resolved −ICOHP values were evaluated along out-of-plane (c) directions. These values provide a quantitative measure of bond covalency and were used to examine correlations with $P_r$ and $E_c$. The bonding analysis was performed on reduced supercells to maintain computational feasibility while preserving local bonding characteristics.

**Acknowledgements**
This study was supported by the Ministry of Education, Culture, Sports, Science and Technology (MEXT) (Nos. 24H00042), and New Energy and Industrial Technology Development Organization (NEDO). PYC would acknowledge the support of JST SPRING (Grant Number JPMJSP2108).

**Data Availability Statement**

# Supporting Information

**Supporting Information for: Decoupling structural and bonding effects on ferroelectric switching in ScAlN via molecular dynamics under an applied electric field**

[1]Ryotaro Sahashi, [1]Po-Yen Chen, and [2]Teruyasu Mizoguchi
[1]Department of Materials Engineering, The University of Tokyo, Tokyo, Japan
[2]Institute of Industrial Science, The University of Tokyo, Tokyo, Japan

This Supporting Information provides validation of the simulation settings used in the molecular dynamics (MD) simulations under an applied electric field calculations, as well as additional structural and bonding analyses supporting the main conclusions of the study.

## S1. Natural Evolution of the Internal Structural Parameter $u$ and Bond Strength

To illustrate the fundamental challenge addressed in this study, we first evaluated the natural composition dependence of the $u$ parameter and the average bond strength in fully relaxed $Sc_xAl_{1-x}N$ models without any artificial constraints. As shown in **Figure S2**, increasing the Sc composition simultaneously induces an increase in the $u$ parameter (indicating structural flattening) and a decrease in the average -ICOHP value (indicating bond weakening). This strong inherent coupling between structural and compositional changes highlights the impossibility of isolating their individual contributions to the ferroelectric properties through conventional compositional variations alone. This intrinsic limitation directly motivates the computational decoupling strategy employed in the main text.

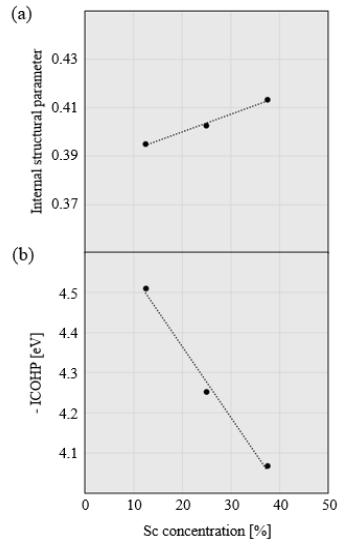

**Figure S1.** Composition dependence of the internal structural parameter $u$ and bond strength in fully relaxed $Sc_xAl_{1-x}N$. (a) The internal parameter $u$ as a function of Sc composition $x$. (b) The average integrated crystal orbital Hamilton population (-ICOHP) as a function of Sc composition $x$. The results clearly demonstrate that structural flattening ($u$ increase) and bond weakening (-ICOHP decrease) occur simultaneously with increasing Sc concentration.

## S2. Structural Control Dataset

To isolate the influence of structural distortion from compositional effects, a structural-control dataset was generated by applying in-plane strain to the $x = 0.25$ composition while keeping the chemical composition fixed. The in-plane lattice constant was systematically varied, and the structures were fully relaxed under the corresponding strain conditions.

As a result, the lattice parameters change continuously with applied strain, leading to a systematic variation of the internal structural parameter $u$, which characterizes the relative displacement of cations and anions along the $c$-axis in the wurtzite lattice. As shown in **Figure S2**, the internal parameter $u$ increases monotonically with tensile in-plane strain and decreases with compressive strain.

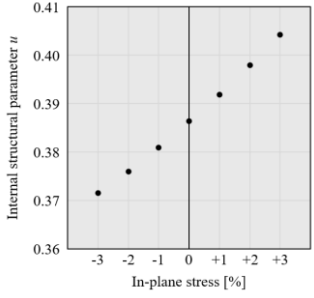

**Figure S2.** Relationship between in-plane strain and internal structural parameter $u$. Variation of lattice parameters and the resulting internal structural parameter $u$ under applied in-plane strain for the $x = 0.25$ composition. The in-plane lattice constant was systematically varied to introduce controlled structural distortion while keeping the chemical composition fixed. This procedure enables continuous tuning of the internal structural parameter $u$ without changing the composition and provides the structural-control dataset used to analyze the structure–property relationships shown in Figure 2 of the main text.

This structural tuning provides a controlled way to modify the atomic configuration associated with polarization switching without introducing changes in chemical bonding arising from composition variation. The resulting dataset therefore serves as a structural reference for evaluating whether the composition dependence of the ferroelectric properties observed in the main text can be explained solely by structural distortion.

**S3. Bonding Analysis**
The bond-strength analysis based on −ICOHP values is summarized in **Figure S3**, which presents correlations between bond strength and the ferroelectric properties $P_r$ and $E_c$. The −ICOHP value is used as an indicator of the cation–anion bond strength in the wurtzite lattice.

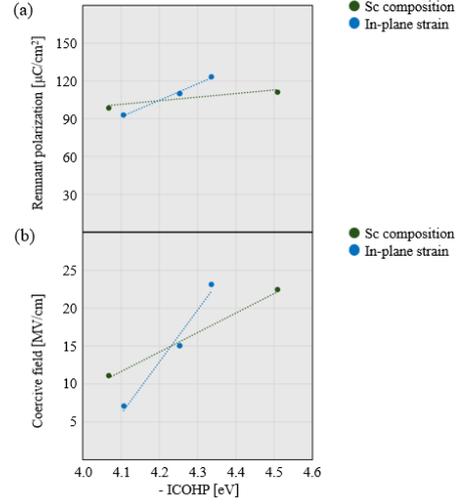

**Figure S3.** Bond strength correlations. (a) Relationship between bond strength (−ICOHP) and $P_r$. (b) Relationship between bond strength (−ICOHP) and $E_c$.

As shown in Figure S3(a), the $P_r$ values obtained from the composition-dependent simulations completely deviate from the linear relationship defined by the pure bonding effect (where the composition is varied while the $u$ parameter is artificially fixed). This stark deviation, combined with the fact that the compositional $P_r$ data perfectly aligns with the structural trend in Figure 2 of the main text, definitively demonstrates that $P_r$ is virtually independent of the chemical bond strength. Instead, the reduction in $P_r$ is governed exclusively by the structural distortion (i.e., the internal structural parameter $u$). In contrast, a more complex behavior is observed for the $E_c$. As shown in Figure S3(b), the composition-dependent $E_c$ data also systematically deviate from the pure bonding trend. As discussed in the main text, the compositional $E_c$ values also deviate from the pure structural relationship (Figure 2). These dual deviations clearly indicate that the composition dependence of the switching field cannot be explained by either the structural effect or the bonding effect alone. Therefore, it is concluded that the reduction in $E_c$ upon Sc incorporation is governed by a superposition of both the structural flattening and the changes in chemical bonding.

## S4. Representative atomic structures along the switching pathway

To illustrate the atomic-scale switching mechanism, representative structures along the nudged elastic band (NEB) pathway are shown in **Figure S4** for the $Sc_{0.25}Al_{0.75}N$ composition calculated using the variable-$u$ model.

framework of the lattice. The intermediate structure corresponds to the highest-energy configuration along the minimum-energy pathway obtained from the NEB calculation. These representative structures provide a visual illustration of the polarization switching process discussed in the main text.

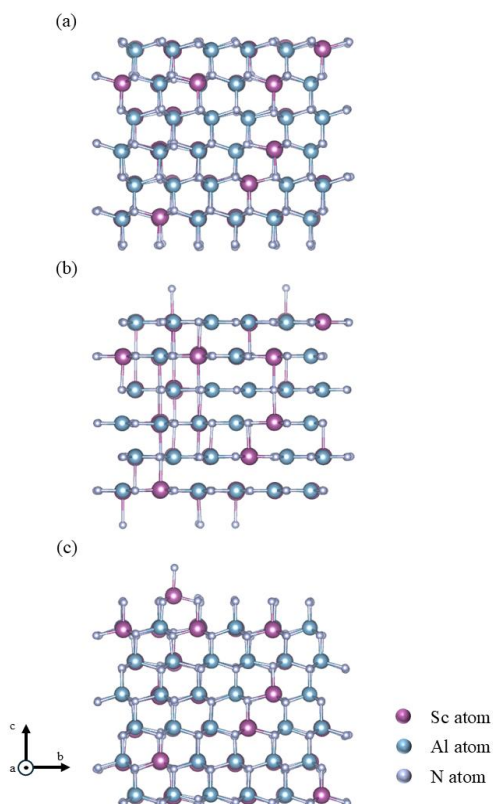

**Figure S4.** Representative atomic structures along the ferroelectric switching pathway obtained from the NEB calculation for $Sc_{0.25}Al_{0.75}N$ using the variable-$u$ model. (a) Initial N-polar ferroelectric state, (b) intermediate structure near the transition state along the minimum energy pathway, and (c) final M-polar switched ferroelectric state. The intermediate structure corresponds to the configuration with the highest energy along the calculated reaction pathway.

The structures correspond to the initial ferroelectric state, the intermediate configuration near the transition state, and the final switched state.

As shown in Figure S4, the switching process proceeds mainly through the displacement of cations along the polar axis, while maintaining the wurtzite-derived

14